\documentclass[journal,12pt,onecolumn]{IEEEtran}
\ifCLASSINFOpdf

\else

\fi

\usepackage{graphicx}
\usepackage{latexsym,bm}
\usepackage{amsmath}
\usepackage{amssymb}
\usepackage{amsthm}
\usepackage{epstopdf}
\usepackage{cite}
\usepackage{mathrsfs}
\usepackage{multirow}
\usepackage[margin=1.0in]{geometry}
\usepackage{booktabs}
\usepackage{array}
\usepackage{pgfplots} 
\usepackage{pbox}
\usepackage{epsfig}
\usepackage{subcaption}
\pgfplotsset{compat=newest} 
\pgfplotsset{plot coordinates/math parser=false}
\usepackage{algorithm}
\usepackage{algpseudocode}
\usepackage{slashbox}
\makeatletter
\def\therule{\makebox[\algorithmicindent][l]{\hspace*{.5em}\vrule height .75\baselineskip depth .25\baselineskip}}%
\newcolumntype{?}{!{\vrule width 1pt}}

\newtoks\therules
\therules={}
\def\appendto#1#2{\expandafter#1\expandafter{\the#1#2}}
\def\gobblefirst#1{
  #1\expandafter\expandafter\expandafter{\expandafter\@gobble\the#1}}%
\def\LState{\State\unskip\the\therules}
\def\pushindent{\appendto\therules\therule}%
\def\popindent{\gobblefirst\therules}%
\def\printindent{\unskip\the\therules}%
\def\printandpush{\printindent\pushindent}%
\def\popandprint{\popindent\printindent}%

\algdef{SE}[WHILE]{While}{EndWhile}[1]
  {\printandpush\algorithmicwhile\ #1\ \algorithmicdo}
  {\popandprint\algorithmicend\ \algorithmicwhile}%
\algdef{SE}[FOR]{For}{EndFor}[1]
  {\printandpush\algorithmicfor\ #1\ \algorithmicdo}
  {\popandprint\algorithmicend\ \algorithmicfor}%
\algdef{S}[FOR]{ForAll}[1]
  {\printindent\algorithmicforall\ #1\ \algorithmicdo}%
\algdef{SE}[LOOP]{Loop}{EndLoop}
  {\printandpush\algorithmicloop}
  {\popandprint\algorithmicend\ \algorithmicloop}%
\algdef{SE}[REPEAT]{Repeat}{Until}
  {\printandpush\algorithmicrepeat}[1]
  {\popandprint\algorithmicuntil\ #1}%
\algdef{SE}[IF]{If}{EndIf}[1]
  {\printandpush\algorithmicif\ #1\ \algorithmicthen}
  {\popandprint\algorithmicend\ \algorithmicif}%
\algdef{C}[IF]{IF}{ElsIf}[1]
  {\popandprint\pushindent\algorithmicelse\ \algorithmicif\ #1\ \algorithmicthen}%
\algdef{Ce}[ELSE]{IF}{Else}{EndIf}
  {\popandprint\pushindent\algorithmicelse}%
\algdef{SE}[PROCEDURE]{Procedure}{EndProcedure}[2]
   {\printandpush\algorithmicprocedure\ \textproc{#1}\ifthenelse{\equal{#2}{}}{}{(#2)}}%
   {\popandprint\algorithmicend\ \algorithmicprocedure}%
\algdef{SE}[FUNCTION]{Function}{EndFunction}[2]
   {\printandpush\algorithmicfunction\ \textproc{#1}\ifthenelse{\equal{#2}{}}{}{(#2)}}%
   {\popandprint\algorithmicend\ \algorithmicfunction}%
\makeatother
\DeclareMathAlphabet{\mathscrbf}{OMS}{mdugm}{b}{n}

\usepackage{array}
\usepackage{array,colortbl,xcolor}

\newcommand\Thickvrule[1]{%
  \multicolumn{1}{!{\vrule width 2pt}c!{\vrule width 2pt}}{#1}%
}

\newlength\Origarrayrulewidth
\newcommand{\Cline}[1]{%
  \noalign{\global\setlength\Origarrayrulewidth{\arrayrulewidth}}%
  \noalign{\global\setlength\arrayrulewidth{2pt}}\cline{#1}%
  \noalign{\global\setlength\arrayrulewidth{\Origarrayrulewidth}}%
}
\usepackage{color}

\begin{document}
\begin{titlepage}
\begin{center}
\vspace*{-2\baselineskip}
\begin{minipage}[l]{7cm}
\flushleft
\includegraphics[width=2 in]{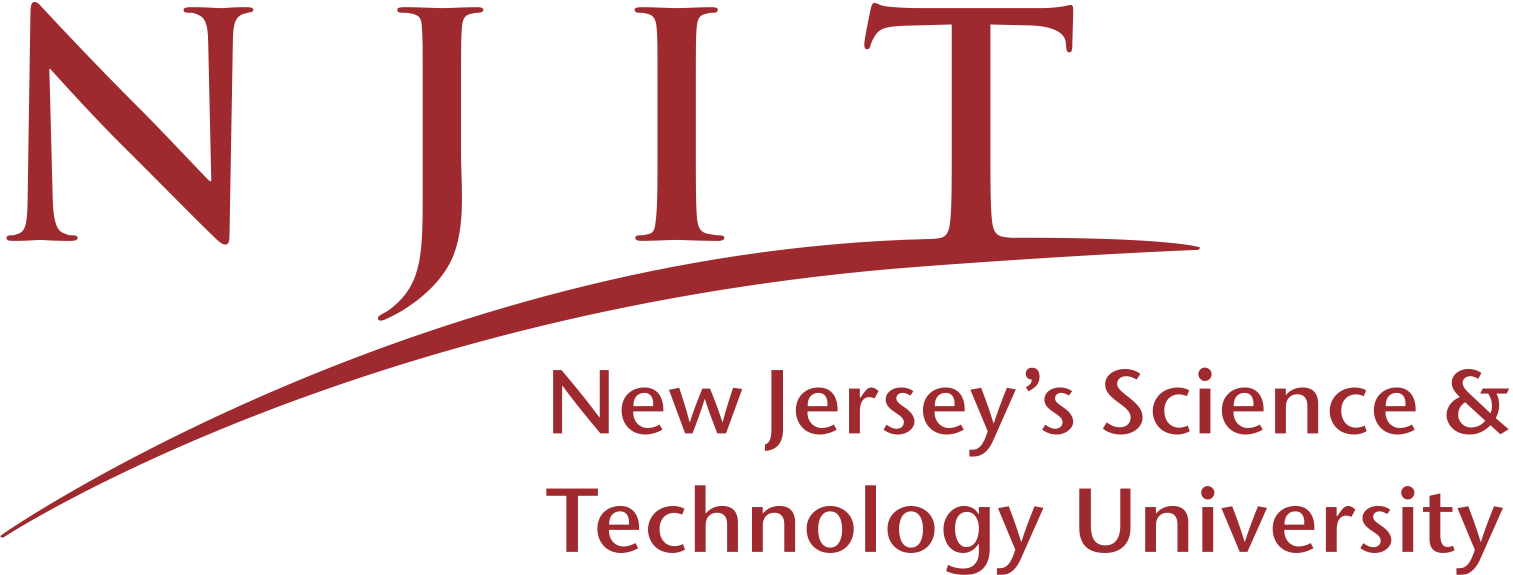}
\end{minipage}
\hfill
\begin{minipage}[r]{7cm}
\flushright
\includegraphics[width=1 in]{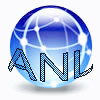}%
\end{minipage}

\vfill

\textsc{\LARGE Content Caching and Distribution in\\ [12pt]
Smart Grid Enabled Wireless Networks}

\vfill
\textsc{\LARGE XUEQING HUANG\\[12pt]
\LARGE NIRWAN ANSARI}\\
\vfill
\textsc{\LARGE TR-ANL-2016-001\\[12pt]
\LARGE March 17, 2016}\\[1.5cm]
\vfill
{ADVANCED NETWORKING LABORATORY\\
 DEPARTMENT OF ELECTRICAL AND COMPUTER ENGINEERING\\
 NEW JERSY INSTITUTE OF TECHNOLOGY}
\end{center}
\end{titlepage}


\begin{abstract}
To facilitate wireless transmission of multimedia content to mobile users, we propose a content caching and distribution framework for smart grid enabled OFDM networks, where each popular multimedia file is coded and distributively stored in multiple energy harvesting enabled serving nodes (SNs), and the green energy distributively harvested by SNs can be shared with each other through the smart grid. The distributive caching, green energy sharing and the on-grid energy backup have improved the reliability and performance of the wireless multimedia downloading process. To minimize the total on-grid power consumption of the whole network, while guaranteeing that each user can retrieve the whole content, the user association scheme is jointly designed with consideration of resource allocation, including subchannel assignment, power allocation and power flow among nodes, where the user association scheme decides which SN serves which user. First, the optimal power allocated to each subchannel and the power flows among nodes are derived analytically. Then, the user association problem is decoupled from the subchannel assignment problem, and the proposed user association scheme has leveraged the multicasting nature of the content download process. To decide which subchannel is assigned to which node, the number of subchannels assigned to each node is first determined, according to the green energy generating rate and the associated traffic offload, and then the corresponding subchannels are selected from the total assigned system spectrum. Simulation results demonstrate that bringing content, green energy and serving node closer to the end user can notably reduce the on-grid energy consumption.
\end{abstract}

\IEEEpeerreviewmaketitle
\section{Introduction}
Wireless content distribution, such as live streaming and on-demand video streaming, is becoming the trending and expected core application for smart phone users. The throughput performance of the wireless access networks and the life time of the battery-operated end devices will be two key factors that limit the user experience. To reduce the user experienced latency, various schemes have been designed to bring the content closer to the users, such as small cell deployment and cloudlet content caching. Given that the probability of a single transmission from the base station is useful for more than one user is rather high for the some of the applications, content reuse schemes have also been designed \cite{7081087,7297864,6336688,7396148}. 

Replication is the simplest cache scheme adopted by many distributed cache systems. For a system with $M$ disttibutive storage nodes, an identical copy of the file is kept at each node, and the whole file can be downloaded from any node \cite{Rodrigues:2005:HAD:2138958.2138987}. The \emph{effective redundancy factor}, which measures the size ratio between the total cached content and the original file, is $M/1$. 

Instead of mirroring the entire content, coded content caching is proposed to reduce the redundancy and improve the storage efficiency. The erasure/regeneration code means that a file is divided into $D$ initial fragments and then coded into $M$ pieces, which will be separately stored in each storage node \cite{5550492,6158613}. The whole file can be retrieved from any $D$ coded pieces, where $D$ is the \emph{reconstruction degree}. The effective redundancy factor of the erasure code is decreased to $M/D$. 

%


By leveraging the ultra-dense deployment of small cells and the advancement of device-to-device communications, popular contents can be cached in different tiers, ranging from macro cells to the device tier. The heterogeneous caching nodes can increase the robustness of the content availability and further support user mobility \cite{Vasilakos:2012:PSN:2342488.2342502}. In addition to the reliability, users can download the desired content without going through backhaul links and retrieve it from core networks. The shorter transmission distance will save radio resources and improve the user experience in terms of traffic latency. Shanmugam. \emph{et al.} \cite{6600983} proposed femtocaching, where coded/uncoded contents are cached in multiple small cells. Given the file requests distribution and the storage size of each femtocell, the placement of the content is studied such that the downloading time is minimized.

In addition to the content storage schemes, the wireless content distribution scheme design focuses on how to deliver the cached content to the ultimate destinations. When multiple users are connected to one serving node, this node can unicast the stored content to each interested user on a dedicated channel with a customized rate depending on its channel conditions, or the node can multicast the content once to the interested users with a multicasting rate that is limited by the worst channel conditions among them. The comparision between unicast and multicast schemes has been conducted in \cite{6516553}.

The storage space and radio resources available at each storage and transmitting node are crucial to the performances of the above mentioned coded content cache and distributive schemes. Although the storage space at a base station or cloudlet is not a big concern because of the advances of hardware technologies and the low cost of hard disk \cite{Chang:2008:BDS:1365815.1365816}, the radio resources needed to transmit and receive data will increase with the data size. To improve the energy efficiency of the wireless content caching and distribution, \emph{energy harvesting} (EH) has been proposed to take fuel from readily available ambient sources, such as wind, solar, and even radio frequency signals \cite{7001217,7143338}. Sharma \emph{et al.} \cite{Sharma:2013:GAO:2483977.2484009} proposed green cache for the off grid EH-enabled base stations. The servers with multimedia caches will blink dynamically from active to inactive states according to the available energy level. Zhou \emph{et al.} \cite{7081087} proposed green delivery for the EH-powered small cells. Proactive caching and push have been jointly designed to resolve the randomness and dynamics of harvested green energy and content requests. 

In this paper, we propose the content caching and distribution for the smart grid enabled heterogeneous networks, where each popular file is coded and stored in multiple EH-enabled serving nodes, and the green energy distributively harvested by these nodes can be shared with each other through the smart grid. The distributive caching, green energy sharing and the on-grid energy backup have improved the reliability and performance of the wireless access networks. To minimize the total on-grid power consumption, the user association scheme and the radio resource allocation including spectrum and power are jointly designed. The simulations results have shown that the proposed caching scheme with low effective redundancy factor can save more on-grid energy consumption when the multimedia downloading rate requirement is very high.

\section{System Model and Problem Formulation}
Consider a downlink wireless network with an eNodeB (eNB) and a group of $K$ users (UE) that are closely located with each other. Suppose all of the UEs are interested in one multimedia file. To bring the content and serving node closer to the users, $M$ serving nodes (SNs) are deployed, where SNs can be small cells or just end devices with the capability of data storage and transmission. The node set in the system $\mathcal{M}=\{1,\cdots,M+1\}$ includes node $1$, i.e., eNB, and SN set $\mathcal{M}'=\{2,\cdots,M+1\}$.

\begin{figure}[t]
\centering
\includegraphics[width=3in]{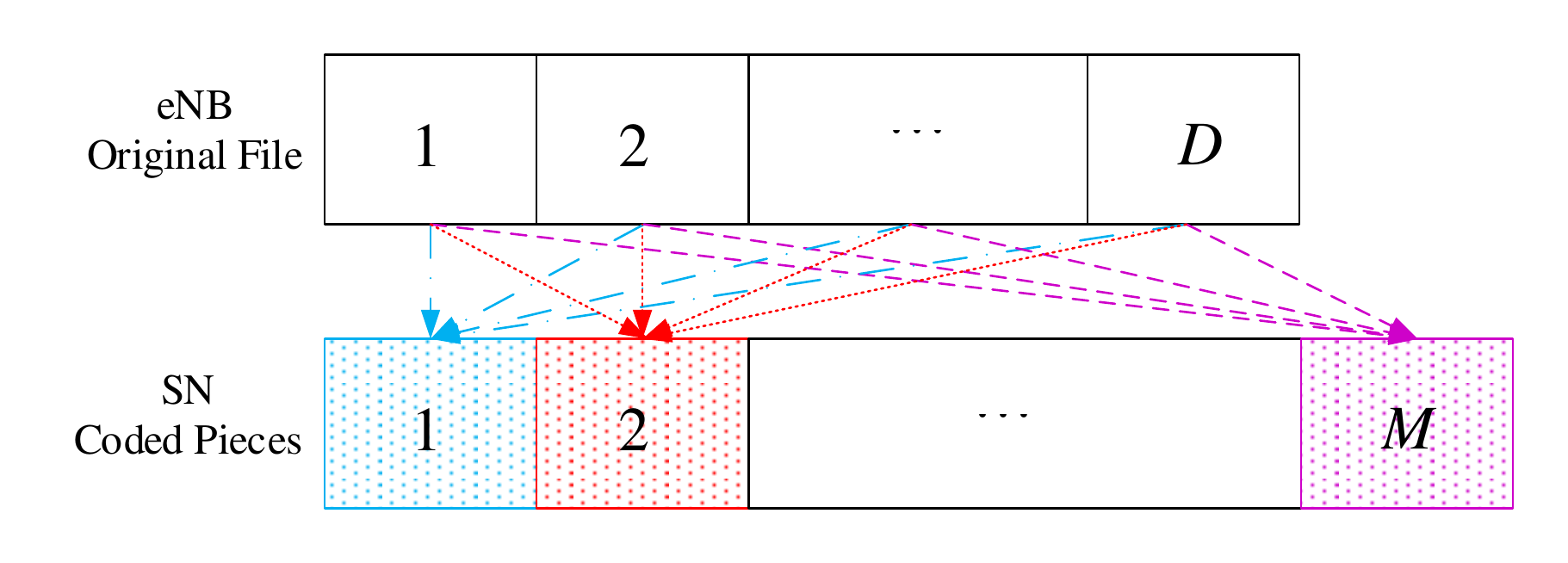}
\caption{Hybrid distributive cached content in each node.}
\label{CONTENT}
\vspace{-1em}
\end{figure}

In particular, as illustrated in Fig. \ref{CONTENT}, eNB maintains one full replica of the multimedia file, and $M$ coded pieces, each of size $S$, are distributively stored in serving nodes, with reconstruction degree $D$. To retrieve the whole file, users can either connect with eNB or any $D$ SNs. 

The total spectrum resources shared by eNB and SNs are divided into $N$ orthogonal subchannels (SCs), each with bandwidth $B$. For Node $i$, define ${\alpha_{i,j}}$ and ${\beta_{i,k}}$ as the indicator functions which will both equal to one when SC $j$ is assigned to the link between node $i$ and UE $k$, and otherwise zero. Then, the subchannel set assigned to node $i$ and the user set that will download the coded content from node $i$ are:
\begin{equation}
\begin{array}{l}
\textit{SC Set: }\mathcal{N}_i=\{j\in\mathcal{N}|\alpha_{i,j}= 1\}\\
\textit{UE Set: }\mathcal{K}_i=\{k\in\mathcal{K}|\beta_{i,k}= 1\}.
\end{array}
\end{equation}

When $\mathcal{K}_i$ consists of more than one user, the $i$-th node will multicast its cached piece to these multiple users, and the multicasting rate is limited by the worst channel conditions among the recipients \cite{6516553}. Denote $P_{i,j}$ as the power allocated by the $i$-th node to subchannel $j$, and the downloading rate of associated node $i$ can be expressed as follows \cite{4432239}:
\begin{equation}
\label{eq1}
R_i=\sum\limits_{j\in\mathcal{N}_i} B\log_2( 1 + \frac{P_{i,j}\min\limits_{k\in\mathcal{K}_i}\{h_{i,j}^k\}}{N_0 B}),\: i\in\mathcal{M},
\end{equation}
where $h_{i,j}^k$ is the channel condition of SC $j$ between node $i$ and UE $k$, which is contributed by the free space path loss which decreases inversely proportional to the square of the distance, the multi path fading, and the shadowing fading which occurs when large objects block paths of propagation. ${N_0}B$ represents the power of additive white Gaussian noise in the $j$-th subchannel.

\begin{figure}[t]
\centering
\includegraphics[width=3in]{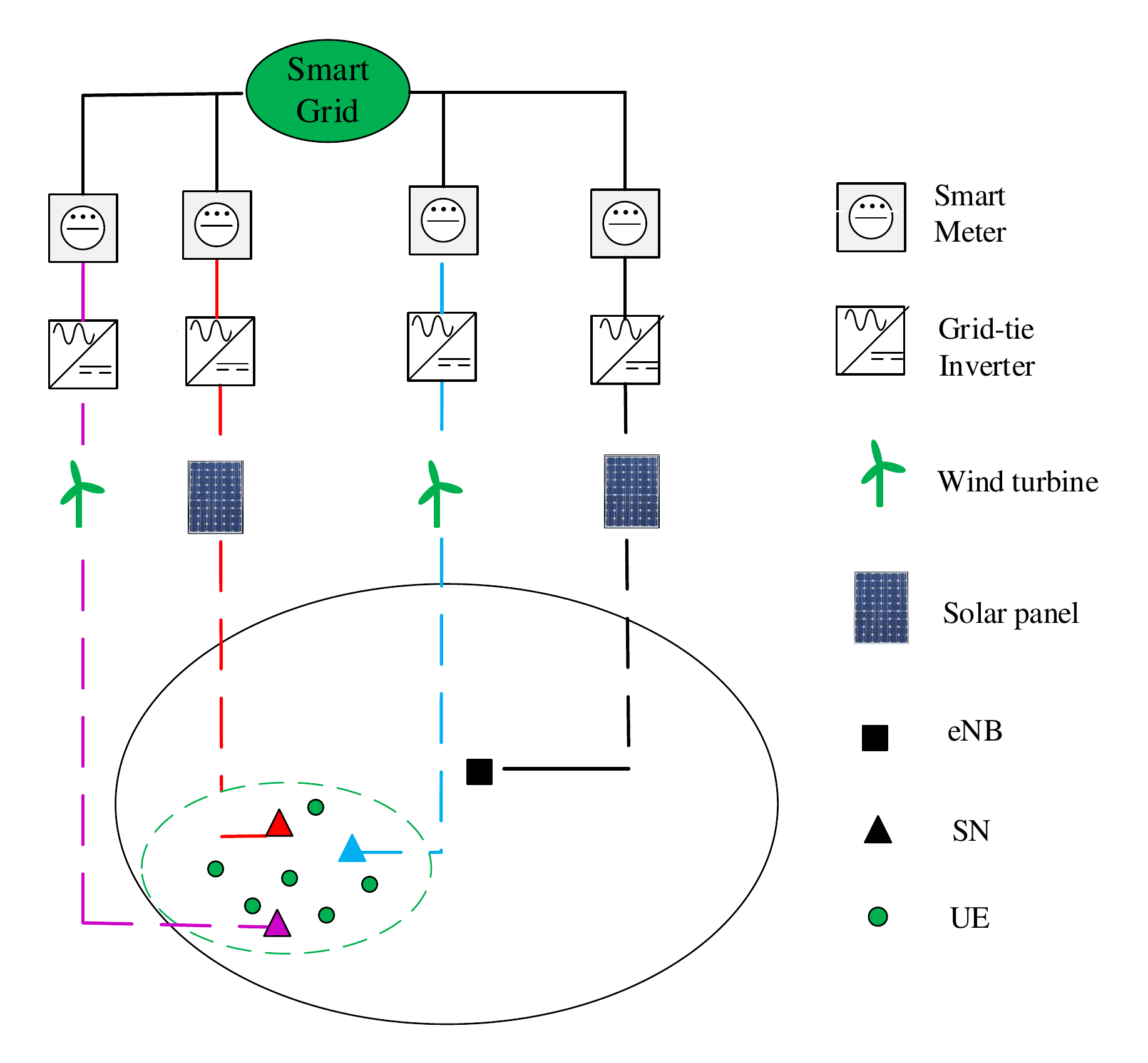}
\caption{Cooperative content distribution system.}
\label{harvesting}
\vspace{-1em}
\end{figure}

All of the nodes in $\mathcal{M}$ are equipped with independent energy harvesters, such as solar panel and wind turbine, as shown in Fig. \ref{harvesting}. Suppose the green energy generating rate of node $i$ is $E_i$, $i\in\mathcal{M}$. In case that the distributively harvested green power is insufficient to support all the data transmission, each node can also withdraw power from the smart grid; let the on-grid power consumed by node $i$ be $P_i^{on}$. Note that if the serving node $i$ is an end user device, $E_i=0$.

In addition to provide on-grid back up power supply, smart grid can be used as the green energy transfer unit between any two nodes. The energy forwarding from one node to another is through the smart grid credit exchange system. Node $i$ can forward $\delta_i$ amount of its own green energy to the grid in exchange for a grid credit. The grid credit can be used by other nodes to purchase on-grid power. In addition, $\theta\le 1$ is the green energy transfer efficiency that is contributed by the power transmission line and the grid-tie inverter, which converts between the harvested DC energy and the transferred AC energy.

Accordingly, the on-grid power consumption of each node is 
\begin{equation}\label{on_grid_PW}
P_i^{on}=\left[\sum\limits_{j \in\mathcal{N}_i} {{P_{i,j}}} - (E_i-\delta_i)\right]^+,\: i\in\mathcal{M},
\end{equation}
where $[\bullet]^+=\max\{\bullet,0\}$. $\delta_i$ denotes the green energy that node $i$ injects into the smart grid ($\delta_i>0$) or the energy that node $i$ draws from the grid using the grid credit ($\delta_i<0$).

Our aim is to minimize the total on-grid power consumption while constructing the original file at all of the UEs. 
\begin{equation}\label{obj}
\begin{array}{l}
\mathop {\min }\limits_{\{\delta_i,{P_{i,j}},{\alpha_{i,j},\beta_{i,k}}\}}  \sum\limits_{i\in\mathcal{M}}{P_i^{on}} \\
\begin{array}{*{20}{l}}
{s.t.}&c_1:&R_i=\left\{\begin{array}{cc}
 DS,&i=1\\
 S, &i\in\mathcal{M'}\\
 \end{array}
 \right.\\
&c_2:&\delta_i\le E_i,i\in\mathcal{M}\\
&c_3:&\sum\limits_{i\in\mathcal{M}}\delta_i (\theta{\bf{1}}_{\{\delta_i>0\}}+{\bf{1}}_{\{\delta_i<0\}})\ge 0\\
&c_4:&\sum\limits_{i\in\mathcal{M}}\alpha_{i,j}=1,j\in\mathcal{N}\\
&c_5:&\beta_{1,k}+\sum\limits_{i\in\mathcal{M'}}\frac{\beta_{i,k}}{D}\ge 1, k\in\mathcal{K}\\
&c_6:&P_{i,j}\ge 0,i\in\mathcal{M},j\in\mathcal{N}\\
&c_7:&{\alpha_{i,j},\beta_{i,k}}\in\{0,1\},i\in\mathcal{M},j\in\mathcal{N},k\in\mathcal{K}\\
\end{array}
\end{array}
\end{equation}
where $c_1$ guarantees the downloading rate of each node so that users can successfully retrieve the content. $c_2$ imposes that the power injected into the grid should be less than the available green energy. For $c_3$, the indicator function ${\bf{1}}_{\{x\}} = 1$ when $x$ is true; otherwise, ${\bf{1}}_{\{x\}} = 0$. $c_4$ indicates that SC $j$ can only be assigned to one node, and $c_5$ indicates that UE $k$ needs to at least connect to eNB or any $D$ nodes. $c_5$ mandates that the withdrawn power should be no greater than the total grid credit.

Note that 1) the assumption of users being able to connect with multiple serving nodes simultaneously is feasible, owing to recent advances of coordinated multi-point transmission/reception (CoMP) technology, which has been proposed to improve the cell edge user performance by allowing mutiple base stations to serve a single user at the same time \cite{7001710}; 2) the orthogonal channel allocation scheme is adopted among all of the serving nodes, because if there is a subchannel overlap between node $i_1$ and $i_2$, users connected to both nodes cannot receive any data because the data transmitted by two nodes are different. Moreover, because all the SNs are deployed to serve a group of UEs that are not far away from each other, if the spectrum is reused, there may be intolerable interference.

\section{Problem Analysis}
To obtain the minimum on-grid power consumption, we first analytically derive the power allocated to each subchannel $P_{i,j}^k$ and the power flow among nodes $\delta_i$. Then, the user association problem is decoupled from the subchannel assignment problem, and the proposed user association scheme has taken advantage of the multicasting nature of the content download process. To decide which subchannel is assigned to which node, the number of SCs assigned to each node is first determined, according to the green energy generating rate and the associated traffic offload, and then the corresponding subchannels are selected from the total assigned system spectrum. 
\subsection{Power Allocation}
Suppose $\mathcal{N}_i$ and $\mathcal{K}_i$ are given, and the objective function in (\ref{obj}) will only be subjected to the data rate requirement in $c1$ and the law of conservation of energy flow in $c2-c3$. Then, the power allocation problem is decoupled from the power flow design problem, because for any $\mathcal{N}_i$ and $\mathcal{K}_i$, to minimize the total on-grid power consumption in (\ref{on_grid_PW}), the power allocation scheme will have to minimize $\sum\nolimits_{j \in\mathcal{N}_i} {{P_{i,j}}}$ while satisfying the rate requirement. Moreover, each node can minimize the power consumption by itself without losing the global optimality.

According to the water-filling algorithm, the optimal power allocation scheme of node $i$ is
\begin{equation}\label{2a}
P_{i,j}=\left[\lambda_{i}-\frac{1}{\gamma_{i,j}^{min}}\right]^+,i\in\mathcal{M},j\in\mathcal{N}_{i},
\end{equation}
where the normalized signal to noise ratio (SNR) is defined as follows:
\begin{equation}\label{SNR}
\gamma_{i,j}^k={h_{i,j}^k/({N_0 B})},
\end{equation}
where the minimum SNR which is decided by the worst channel conditions among users in $\mathcal{K}_i$, i.e.,
\begin{equation}\label{minSNR}
\gamma_{i,j}^{min}=\min\limits_{k\in\mathcal{K}_i}\{\gamma_{i,j}^k\}.
\end{equation}
The Lagrangian multiplier $\lambda_{i}$ is given as follows:

\begin{equation}\label{2a1}
\lambda_{i}=\left\{\begin{array}{l}
{\left({\frac{2^{\frac{DS}{B}}}{\prod\limits_{j\in\mathcal{N}_{i}}{\gamma ^{min}_{i,j}}}}\right)}^{\frac{1}{{N}_{i}}},\: i=1\\
{\left({\frac{2^{\frac{S}{B}}}{\prod\limits_{j\in\mathcal{N}_{i}}{\gamma ^{min}_{i,j}}}}\right)}^{\frac{1}{{N}_{i}}},\: i\in\mathcal{M'}\\
\end{array}
\right.
\end{equation}
where $N_i=|\mathcal{N}_i|$ is the cardinality of $\mathcal{N}_i$.
\subsection{Power Flow}
Let $\mathcal{M}^{+}=\{i\in\mathcal{M}|\delta_i>0\}$ be the set of nodes which contribute green energy to other nodes, and $\mathcal{M}^{-}=\{i\in\mathcal{M}|\delta_i\le 0\}$ be the set of nodes, which tap on the grid credit. Then,  as stated in the following Lemma, nodes in $\mathcal{M}^+$ will transfer all the residual energy to other nodes through the power grid.

\emph{Lemma 1:} The optimal power flow from node $i$, $i\in\mathcal{M}^+$, to the grid is 
\begin{equation}\label{alpha}
\delta_i=E_i-\sum\limits_{j\in\mathcal{N}_i}P_{i,j}
\end{equation}
\begin{proof}
Since $\delta_i$ in $\mathcal{M}^+$ is greater than zero, (\ref{alpha}) and $c_4$ in (\ref{obj}) imply the following two constraints imposed on $P_{i,j}$ and $P_i^{on}$:
\begin{equation}\label{ana1}
\left\{\begin{array}{l}
\sum\limits_{j\in\mathcal{N}_i}P_{i,j}<E_i,i\in\mathcal{M}_{+}\\
P_i^{on}=\left[\sum\limits_{j \in\mathcal{N}_i} {{P_{i,j}}} - (E_i-\delta_i)\right]^+=0,i\in\mathcal{M}^+.\\
\end{array}\right.
\end{equation}

The intuition behind (\ref{alpha}) is that 1) if $\delta_i^*<\delta_i$ is optimal, the on-grid power consumption of node $i$ remains at zero, but the gird credit will decrease by $\theta(\delta_i-\delta_i^*)$; 2) if $\delta_i^*>\delta_i$ is optimal, the increment in $P_i^{on}$ is $\delta_i^*-\delta_i$, which is greater than the increment in the earned power grid $\theta(\delta_i^*-\delta_i)$. So, in both cases, the total on-grid power consumption of all nodes in $\mathcal{M}$ will increase. 
\end{proof}

According to (\ref{ana1}), the objective function in (\ref{obj}) becomes 
\begin{equation}\label{simpli}
\sum\limits_{i\in\mathcal{M}}{P_i^{on}}=\sum\limits_{i\in\mathcal{M}^-}{P_i^{on}}.
\end{equation}
Moreover, $\delta_i$, $i\in\mathcal{M}^-$, is given as follows.

\emph{Lemma 2:} The optimal grid credit used by node $i$, $i\in\mathcal{M}^-$, is 
\begin{equation}\label{lem2}
\delta_i=-\min\{\sum\limits_{j\in\mathcal{N}_i}P_{i,j}-E_i,\sum\limits_{i'\in\mathcal{M}^+}\theta\delta_{i'}+\sum\limits_{i'\in\mathcal{M}^-\backslash{i}}\delta_{i'}\}
\end{equation}
\begin{proof}
According to $c_6$ in (\ref{obj}), the total grid credit is $\sum\nolimits_{i'\in\mathcal{M}^+}\theta\delta_{i'}$. So, the grid credit available to a certain node $i$, $i\in\mathcal{M}^-$, is $\sum\nolimits_{i'\in\mathcal{M}^+}\theta\delta_{i'}+\sum\nolimits_{i'\in\mathcal{M}^-\backslash{i}}\delta_{i'}$. 

Moreover, since the minimum on-grid power consumption $\left[\sum\nolimits_{j \in\mathcal{N}_i} {{P_{i,j}}} - (E_i-\delta_i)\right]^+$ is zero, the maximum grid credit node $i$ needs is $\sum\nolimits_{j\in\mathcal{N}_i}P_{i,j}-E_i$.

As a result, the optimal grid credit is the minimum value of the available credit and the maximum needed credit.
\end{proof}


According to (\ref{lem2}), we have $\sum\nolimits_{j \in\mathcal{N}_i} {{P_{i,j}}} - (E_i-\delta_i)\ge 0$, and (\ref{simpli}) is further simplified as follows.
\begin{equation}\label{red1}
\sum\limits_{i\in\mathcal{M}}{P_i^{on}}=\sum\limits_{i\in\mathcal{M}^-}[{\sum\limits_{i\in\mathcal{N}_i}P_{i,j}}-(E_i-\delta_i)].
\end{equation}

 Based on the optimal $\delta_i$ given in \emph{Lemmas} 1 ($i\in\mathcal{M}^+$) and 2 ($i\in\mathcal{M}^-$), $\sum\nolimits_{i\in\mathcal{M}^-}\delta_i$ becomes $-min\{\sum\nolimits_{i\in\mathcal{M}^-}(\sum\nolimits_{j\in\mathcal{N}_i}P_{i,j}-E_i),\sum\nolimits_{i\in\mathcal{M}^+}\theta\delta_{i}\}$, and the objective function in (\ref{red1}) can be further reduced.
\begin{equation}\label{red}
\begin{array}{l}
\sum\limits_{i\in\mathcal{M}}{P_i^{on}}=\sum\limits_{i\in\mathcal{M}^-}({\sum\limits_{j\in\mathcal{N}_i}P_{i,j}}- E_i)+\sum\limits_{i\in\mathcal{M}^-}\delta_i=\\
\sum\limits_{i\in\mathcal{M}^-}({\sum\limits_{j\in\mathcal{N}_i}P_{i,j}}- E_i)-\\
min\{\sum\limits_{i\in\mathcal{M}^-}(\sum\limits_{j\in\mathcal{N}_i}P_{i,j}-E_i),\sum\limits_{i\in\mathcal{M}^+}\theta\delta_{i}\}=\\
\max\{0,\sum\limits_{i\in\mathcal{M}^-}{\sum\limits_{j\in\mathcal{N}_i}P_{i,j}}- \sum\limits_{i\in\mathcal{M}^-}E_i-\sum\limits_{i\in\mathcal{M}^+}\theta\delta_{i}\}=\\
\max\{0,\sum\limits_{i\in\mathcal{M}^-}{\sum\limits_{j\in\mathcal{N}_i}P_{i,j}}+\sum\limits_{i\in\mathcal{M}^+}{\sum\limits_{j\in\mathcal{N}_i}\theta P_{i,j}}- \\
\sum\limits_{i\in\mathcal{M}^-}E_i- \sum\limits_{i\in\mathcal{M}^+}\theta E_i\}.
\end{array}
\end{equation}


Then, the original problem becomes
\begin{equation}\label{obj1}
\begin{array}{l}
\mathop {\min }\limits_{\{{\alpha_{i,j},\beta_{i,k}}\}}  \sum\limits_{i\in\mathcal{M}^-}{\sum\limits_{j\in\mathcal{N}_i}P_{i,j}} + \sum\limits_{i\in\mathcal{M}^+}{\sum\limits_{j\in\mathcal{N}_i}\theta P_{i,j}}-\\
\sum\limits_{i\in\mathcal{M}^-}E_i- \sum\limits_{i\in\mathcal{M}^+}\theta E_i\\
\begin{array}{*{20}{l}}
{s.t.}&\mathcal{M}^-=\{i\in\mathcal{M}|\sum\limits_{j\in\mathcal{N}_i}P_{i,j}\ge E_i\}\\
&\mathcal{M}^+=\{i\in\mathcal{M}|\sum\limits_{j\in\mathcal{N}_i}P_{i,j}< E_i\}\\
&\sum\limits_{i\in\mathcal{M}}\alpha_{i,j}=1,j\in\mathcal{N}\\
&\beta_{1,k}+\sum\limits_{j\in\mathcal{M'}}\frac{\beta_{i,k}}{D}\ge 1, k\in\mathcal{K}\\
&{\alpha_{i,j},\beta_{i,k}}\in\{0,1\},i\in\mathcal{M},j\in\mathcal{N},k\in\mathcal{K},\\
\end{array}
\end{array}
\end{equation}
where $P_{i,j}$ is given in (\ref{2a}).

The optimization problem in (\ref{obj1}) is the integer programming problem, which is NP-hard in general \cite{Papadimitriou:1981:CIP:322276.322287}. The proof of NP-hardness is omitted due to the page limit; detailed proofs for similar subchannel and power allocation problems are referred to \cite{5506244,6336839,6884803}.
\subsection{Decoupling between User Association and Subchannel Assignment}\label{assump}
To study the characteristics of the above mentioned NP-hard problem, we first make the following assumptions which will relax the problem.

{\bf{Assumption 1}}: For the link between node $i$ and user $k$, the average value of SNR across all of the $N$ subchannels will determine the user association scheme and subchannel assignment scheme.

Define $\gamma_{i}^k$ as the average value of SNR across all of the $N$ subchannels:
\begin{equation}
\gamma_{i}^k={\sum\limits_{j\in\mathcal{N}}\gamma_{i,j}^k}/{N}.
\end{equation}
The design of $\gamma_{i}^k$ is consistent with the current OFDM based wireless access system, where the serving node is selected based on the reference signal receive power (RSRP), which is the average power of reference signals (RSs) over the entire bandwidth \cite{1244793}. 

Consequently, the subchannel assignment problem becomes choosing the number of SCs that are allocated to node $i$.

{\bf{Assumption 2}}: $E_i=0$, $i\in\mathcal{M}$.

As we can see from the objective function in (\ref{obj1}), for any given $\mathcal{M}^+$ and $\mathcal{M}^-$, minimizing the on-grid power consumption of all nodes means minimizing the total power consumption of all nodes, and the power consumption of nodes in set $\mathcal{M}^+$ has an additional discount factor $\theta$. Based on this observation, $\mathcal{M}^-$ in (\ref{obj1}) becomes the whole node set $\mathcal{M}$, because the above implies that none of the nodes can harvest energy from the green energy sources.

With the above two assumptions and the optimal power allocation scheme derived in (\ref{2a}), the objective function becomes minimizing the total power consumption of all nodes.
\begin{equation}\label{obj1_2}
\begin{array}{l}
\mathop {\min }\limits_{\{N_i,\beta_{i,k}\}}  \frac{N_1(2^\frac{DS}{N_1B}-1)}{\gamma_1^{min}}+\sum\limits_{i\in\mathcal{M}'}\frac{N_i(2^\frac{S}{N_iB}-1)}{\gamma_i^{min}} \\
\begin{array}{*{20}{l}}
{s.t.}& \sum\limits_{i\in\mathcal{M}}N_i=N\\
&\beta_{1,k}+\sum\limits_{i\in\mathcal{M'}}\frac{\beta_{i,k}}{D}\ge 1, k\in\mathcal{K},\\
\end{array}
\end{array}
\end{equation}
where $\beta_{i,k}$ determines the minimum average SNR of each node $\gamma_i^{min}$:
\begin{equation}
\gamma_i^{min}=\min\limits_{k\in\mathcal{K}_i}\{\gamma_{i}^k\}.
\end{equation}

In addition to the assigned subchannel number $N_i$ and the minimum average SNR of each node $\gamma_i^{min}$, an implicit variable that will impact the total power consumption is the total traffic load requested in the system. Define $D^*$ as the number of 
coded/uncoded file fragments to be transmitted to the users. The corresponding traffic load $D^*S$ is classified into the following categories.

\emph{Case I:} all of the UEs are assigned to eNB, then $\mathcal{K}_1=\mathcal{K}$ and $\mathcal{N}_1=\mathcal{N}$. The traffic load is $DS$, i.e.,
 \[D^*=D.\]

\emph{Case II:} all of the UEs are assigned to SNs, and then at least $D$ SNs of the total $M$ nodes in $\mathcal{M}'$ are active, where node $i$ being active means node $i$ needs to transmit data to at least one user. The traffic load is between $DS$ and $MS$, i.e.,
 \[D \le D^*\le M.\]

\emph{Case III:} both eNB and SNs are active, and the traffic load is between $2DS$ and $(M+D)S$, i.e.,
 \[2D \le D^*\le M+D.\]

In the following, we decouple the user association scheme $\beta_{i,k}$ from subchannel assignment scheme $N_i$. By solving the two subproblems separately, a good, but not necessarily optimal, solution is found that limits a certain level of on-grid power consumption for the system.

\subsection{User Association}
As we can see, \emph{Case I} can be chosen as the baseline, because 1) it yields the minimum traffic load, and 2) the total spectrum can be considered proportionally allocated to the assigned traffic load. To offload the traffic to serving nodes with local cache, we have to make sure that the increased minimum average SNR ${\gamma_i^{min}}$ will compensate for the increased traffic load $D^*$, in terms of saving total power consumption. With $N_1=ND/D^*$ and $N_i=N/D^*$, the trade-off between the increased traffic load $D^*$ and increased ${\gamma_i^{min}}$ is given below.

\begin{equation}\label{obj1_3}
\begin{array}{l}
\mathop {\min }\limits_{\{\beta_{i,k}\}} \frac{N}{D^*}(2^{\frac{D^*S}{NB}}-1)(\frac{D}{\gamma_1^{min}}+\sum\limits_{i\in\mathcal{M}^*}\frac{1}{\gamma_i^{min}}) \\
\begin{array}{*{20}{l}}
{s.t.}&\beta_{1,k}+\sum\limits_{i\in\mathcal{M'}}\frac{\beta_{i,k}}{D}\ge 1, k\in\mathcal{K}\\
\end{array}
\end{array}
\end{equation}
where $\mathcal{M}^*\in\mathcal{M}'$ is the set of active serving nodes, and $|\mathcal{M}^*|=D^*-D$.\

As we can see, the total power consumption in (\ref{obj1_3}) will increase exponentially with $D^*$, while decrease with ${\gamma_1^{min}}$. So we iterate $D^*$ from $D$ to $D+M$, and in each iteration, we use the following three steps to find the user association scheme, that will yield as high $\gamma_i^{min}$ as possible. 

In the first step, $\mathcal{M}_k$, the set of nodes which store $D$ coded/uncoded fragments that are closest to user $k$, is found. In the second step, the backup nodes will be found if the user will not degrade the $\gamma_i^{min}$ that is given in the first step. In the third step, if a user $k$ has a backup node, and this user is the farthest away from node $i$ that is picked in the first step, i.e., $\gamma_i^k=\gamma_i^{min}$, then node $i$ will be dropped from $\mathcal{M}_k$ and the backup node is chosen.

The intuition for the first step is that users should be connected to the most closely located serving nodes so that the reconstriuction requirement is satisfied, and the minimum $\gamma_i^k$ is high. The intuition for the second step and third step is owing to the multicasting nature of the link between node $i$ and users in $\mathcal{K}_i$; the ultimate factor that dictates the energy consumption is $\gamma_i^{min}$. Consequently, even when node $i_1$ is closer to UE $k$ than node $i_2$, if $\gamma_{i_1}^{min}=\gamma_{i_1}^k$ and $\gamma_{i_2}^{min}<\gamma_{i_2}^k$, user $k$ will be choose node $i_2$ over $i_1$ as the downloading node. In this case, $\gamma_{i_1}^{min}$ will increase, while $\gamma_{i_2}^{min}$ remains the same, and the total power consumption will be reduced.

The user association algorithm that follows the above mentioned three steps is given in Alg. \ref{euclid}.

\begin{algorithm}
\caption{Maximal minimum SNR user association algorithm}\label{euclid}
\begin{algorithmic}[1]
\LState {$\mathcal{M}_k=\emptyset,k\in\mathcal{K}$}\Comment{Node set for UE $k$}
\For {$k\in\mathcal{K}$}\Comment {Step 1}
\LState {Sort nodes in $\mathcal{M}$ in the descending order of $\gamma_{i}^k$}
\LState {Index node $i$ with the $IX_i^k$-th largest SNR to UE $k$}
\If {$IX_{1}^k\le D$}
\LState {$\mathcal{M}_k=\{i\in\mathcal{M}|IX_{i}^k\le D+1\}$}
\If {$\sum\limits_{i\in\mathcal{M}_k\backslash\{1\}}\gamma_{i}^k>D\gamma_{1}^k$}
\LState {$\mathcal{M}_k=\mathcal{M}_k\backslash\{1\}$}
\Else
\LState {$\mathcal{M}_k=\{1\}$}
\EndIf
\Else
\LState {$\mathcal{M}_k=\{i\in\mathcal{M}|IX_{i}^k\le D\}$}
\EndIf
\EndFor
\For {$k\in\mathcal{K}$}\Comment {Step 2}
\For {$i\in\mathcal{M}_k$}
\For {$k'\in\mathcal{K}\backslash{k}$}
\If {$\gamma_{i}^{k'}>\gamma_{i}^k$}
\LState {$\mathcal{M}_{k'}=\mathcal{M}_{k'}\cup\{i\}$}
\EndIf
\EndFor
\EndFor
\EndFor

\LState {$\beta_{i,k}=0,i\in\mathcal{M},k\in\mathcal{K}$}\Comment {Step 3}
\LState {$\mathcal{K}_i=\emptyset,i\in\mathcal{M}$}
\LState {Update $\beta_{i,k}$, $\mathcal{K}_i$, and $\gamma_i^{min}$ according to $\mathcal{M}_k$}
\LState {$\mathcal{K}^*=\left\{k\in\mathcal{K}\left|\begin{array}{l}\gamma_{i}^k=\gamma_i^{min},\exists i\in\mathcal{M}_k\\
\beta_{1,k}+\sum\limits_{i\in\mathcal{M}'}\frac{\beta_{i,k}}{D}>1\\
\end{array}\right.\right\}$}
\While {$\mathcal{K}^*\neq\emptyset$}
\For {$k\in\mathcal{K}^*$}
\LState {$\mathcal{M}_k^*=\{i\in\mathcal{M}_k|\gamma_{i}^k=\gamma_i^{min}\}$}
\LState {$i^*=arg \max\limits_{i\in\mathcal{M}_k^*} IX_i^k$}
\If {$i^*= 1$ $\&\&$ $\sum\limits_{i\in\mathcal{M}'}\frac{\beta_{i,k}}{D}<1$}
\LState {$\mathcal{M}_{k}=\{i^*\}$}
\Else
\LState {$\mathcal{M}_{k}=\mathcal{M}_{k}\backslash\{i^*\}$}
\EndIf
\LState {Update $\beta_{i,k}$, $\mathcal{K}_i$, $\gamma_i^{min}$, and $\mathcal{M}_k^*$ according to $\mathcal{M}_k$}
\EndFor
\LState {Update $\mathcal{K}^*$}
\EndWhile
\LState {$\mathcal{M}_{act}=\{i\in\mathcal{M}|\mathcal{K}_i\neq\emptyset\}$}\Comment{Node set which has at least one user}
\LState\Return $\mathcal{M}_{act}$, $\beta_{i,k}$, $\mathcal{K}_i$, $i\in\mathcal{M},k\in\mathcal{K}$
\end{algorithmic}
\end{algorithm}


The user association scheme is designed based upon {\bf{Assumption 2}} in Section \ref{assump}, which means that neither eNB nor SNs can harvest energy. With positive $E_i$, however, some nodes which have sufficiently large green energy may belong to $\mathcal{M}^+$. Since the total power consumption of these nodes have a discount factor $\theta$, traffic offloading will occur when the following constraint is satisfied.
\begin{equation}\label{offload}
P_{i_1}-P'_{i_1}>(P'_{i_2}-P_{i_2})\theta,
\end{equation}
where $i_1\in\mathcal{M}^-$ and $i_2\in\mathcal{M}^+$. $P_{i_1}$ and $P_{i_2}$ are the total power consumption of node $i_1$ and node $i_2$ before traffic offloading, respectively. $P'_{i_1}$ and $P'_{i_2}$ are the corresponding power consumption after traffic offloading.
The left hand is the energy that is saved by node $i_1$, and the right hand is the increment of energy consumption of node $i_2$. 

Instead of designing the traffic offloading scheme and changing the user association result, we propose to manipulate the following subchannel assignment scheme, such that node $i_1$ will get more spectrum resources, and node $i_2$ will be assigned less subchannels.

\subsection{Subchannel Assignment}
Given the user association scheme $\beta_{i,k}$ in Alg. \ref{euclid}, the minimum average SNR $\gamma_{i}^{min}$ becomes known. Then, without {\bf{Assumption 2}} in Section \ref{assump}, the problem becomes finding the optimal $N_i$ such that $\sum\nolimits_{i\in\mathcal{M}^-}{\sum\nolimits_{i\in\mathcal{N}_i}P_{i,j}} + \sum\nolimits_{i\in\mathcal{M}^+}{\sum\nolimits_{i\in\mathcal{N}_i}\theta P_{i,j}}-\sum\nolimits_{i\in\mathcal{M}^-} E_i-\sum\nolimits_{i\in\mathcal{M}^+} \theta E_i$ is minimized.


Denote $P_i^{N_i}$ as the total power consumption of node $i$ when it is assigned with $N_i$ SCs. According to the optimal power allocation in (\ref{2a}), $P_i^{N_i}$ is given as follows. 
\begin{equation}
P_i^{N_i}=\left\{
\begin{array}{*{20}{c}}
\frac{N_i({2^{\frac{DS}{BN_i}}-1})}{\gamma_i^{min}},\:i=1\\
\frac{N_i({2^{\frac{S}{BN_i}}-1})}{\gamma_i^{min}},\:i\in\mathcal{M}'.\\
\end{array}\right.
\end{equation}

Since each active node should at least has one SC, the maximum number of SCs that a node can get is $N-M_{act}+1$, where $M_{act}=|\mathcal{M}_{act}|$ is the number of serving nodes with at least one user, and the serving nodes in $\mathcal{M}_{act}$ returned by Alg. \ref{euclid} will provide the $D^*S$ amount of downloading traffic. Then, we have the following table with $P_i^{N_i}$, $i\in\mathcal{M}$, $N_i\in\{1,\cdots,N-M_{act}+1\}$.

\renewcommand{\arraystretch}{2}
\begin{table}[ht]
\caption{Power Consumption Table} 
\centering\  
\begin{tabular}{|c|c|c|c|c} 
\cline{1-4}                       
 \backslashbox{BS\kern-1em}{\kern-1em SC}   &1  &	$\cdots$	    &${N-M_{act}+1}$ &  \\\hline
{ $1$}   &$P_1^1$  &	$\cdots$	    &$P_1^{N-M_{act}+1}$ &  \Thickvrule{\cellcolor{blue!25}{$P_1^{N_1}-E_1$}} \\\cline{1-4}\Cline{5-5} 
$\vdots$ &$\vdots$  &	$\ddots$	&$\vdots$      &  \Thickvrule{{$\vdots$}}\\\cline{1-4}\Cline{5-5} 
 $i$   &$P_i^1$  &	$\cdots$	    &$P_i^{N-M_{act}+1}$ &  \Thickvrule{\cellcolor{blue!25}{$P_i^{N_i}-E_i$}} \\\cline{1-4}\Cline{5-5} 
$\vdots$ &$\vdots$  &	$\ddots$	&$\vdots$      &  \Thickvrule{{$\vdots$}} \\\cline{1-4}\Cline{5-5} 
 $M+1$   &$P_{M+1}^1$  &	$\cdots$	    &$P_{M+1}^{N-M_{act}+1}$ &  \Thickvrule{\cellcolor{blue!25}{$P_{M+1}^{N_M}-E_M$}} \\\cline{1-4}\Cline{5-5} 
\end{tabular}
\label{tab_1}
\end{table}

To minimize $\sum\nolimits_{i\in\mathcal{M}^-}({P_{i}^{N_i}}-E_i) + \sum\nolimits_{i\in\mathcal{M}^+}\theta ({P_{i}^{N_i}}-E_i)$, where the total available SC is limited by $N$, we use the subchannel allocation algorithm in Alg. \ref{SC}. First, we assume each node is allocated with only one SC. Then, the node with the highest energy consumption will get an additional SC, until all of the $N$ SCs are assigned.
\begin{algorithm}
\newcommand{\Break}{\State \textbf{break} }
\caption{Subchannel allocation algorithm}\label{SC}
\begin{algorithmic}[1]
\LState {$N_i=\left\{\begin{array}{l}
1,i\in\mathcal{M}_{act}\\
0,i\notin\mathcal{M}_{act}\\
\end{array}\right.$}
\LState {Calculate Table I}
\LState $N_{tot}=N-M_{act}+1$
\While {$N_{tot}>0$}
\LState {$i^*=\arg \max\limits_{i\in\mathcal{M}_{act}} P_i^{N_i}-E_i$}
\LState {$N_{i^*}=N_{i^*}+1$}
\LState {Update the last column of Table \ref{tab_1}}
\LState $N_{tot}=N_{tot}-1$
\EndWhile

\LState\Return $N_i$, $i\in\mathcal{M}$
\end{algorithmic}
\end{algorithm}

With the minimum SNR $\gamma_{i,j}^{min}$, $i\in\mathcal{M}$, $j\in\mathcal{N}$ in (\ref{minSNR}), and the number of subchannels assigned to each node provided by Alg. \ref{euclid} and Alg. \ref{SC}, respectively, we can lift {\bf{Assumption 1}} proposed in Section \ref{assump}. The subchannel selection algorithm yields the highest priority to the node with the largest on-grid energy consumption. That is, node $i^*$ in (\ref{alg_eq}) will select the $\mathcal{N}_i$ SCs with the highest SNR.

\begin{algorithm}
\newcommand{\Break}{\State \textbf{break} }
\caption{Subchannel selection algorithm}\label{SC1}
\begin{algorithmic}[1]
\LState {$\mathcal{N}=\{1,\cdots,N\}$}
\While {$\mathcal{M}_{act}\neq \emptyset$}
\LState {\begin{equation}\label{alg_eq}
i^*=\arg \max\limits_{i\in\mathcal{M}_{act}} P_i^{N_i}-E_i
\end{equation}
}
\LState {Sort SCs in $\mathcal{N}$ in the descending order of $\gamma_{i^*,j}$}
\LState {Index SC with the $IX_j$-th largest SNR to node $i^*$}
\LState {$\mathcal{N}_{i^*}=\{j\in\mathcal{N}|IX_j\ge N_i\}$}
\LState {$\alpha_{i^*,j}=1,j\in\mathcal{N}_{i^*}$}
\LState {$\mathcal{N}=\mathcal{N}\backslash\{\mathcal{N}_{i^*}\}$}
\LState {$\mathcal{M}_{act}=\mathcal{M}_{act}\backslash\{i^*\}$}
\EndWhile
\LState {$\mathcal{N}=\{1,\cdots,N\}$}
\LState\Return $\alpha_{i,j}$, $\mathcal{N}_i$, $i\in\mathcal{M}$, $j\in\mathcal{N}$
\end{algorithmic}
\end{algorithm}

So far, we have proposed separate algorithms for four sets of variables in (\ref{obj}). To get the optimal solution, the maximal minimum SNR algorithm in Alg. \ref{euclid} is first adopted to obtain the user association scheme, and then the subchannel assignment scheme in Alg. \ref{SC} and Alg. \ref{SC1} are employed to assign the spectrum to each node. Then, the power is allocated according to (\ref{2a}), and subsequently the power flows given in \emph{Lemmas} 1-2 will yield the minimum on-grid power power consumption.

\section{Simulation Results}
In this section, we assume that the average solar green energy density is $15$ $mW/cm^2$. Different sizes of solar panels equipped in eNB and serving nodes result in diverse average green energy generating rate. For time duration of $T=1$ $ms$, the green energy generating rate for eNB is $E_1=20$ $W$, and for SNs, $E_i=0.5$ $W$, $i\in\mathcal{M}'$. 

\begin{figure}[t]
\centering
\includegraphics[width=3in]{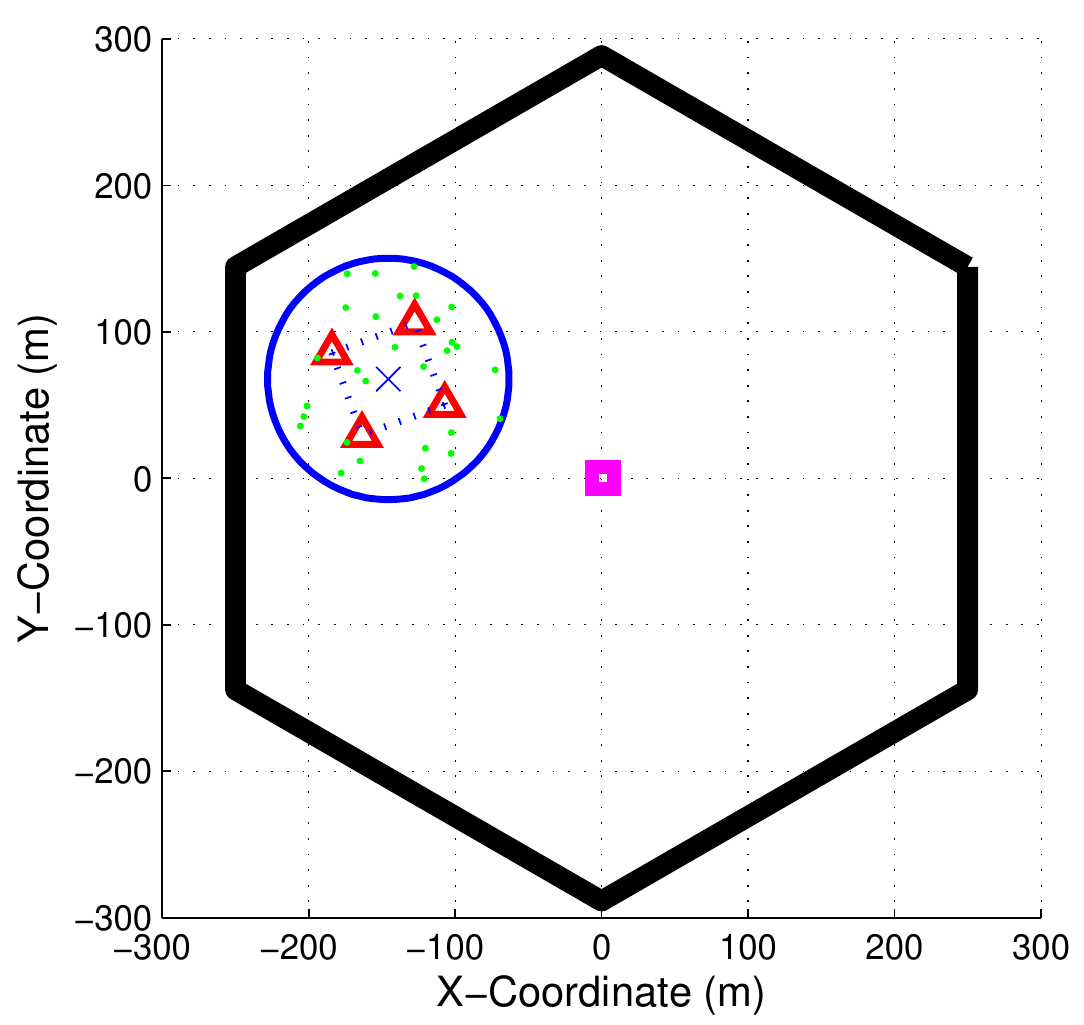}
\caption{Illustration of the system deployment (Triangle: SN; Cross: Virtual SN center; Dot: UE; Square: eNB).}
\label{SCENARIO_PLOT}
\end{figure}

The four SNs form a square area with $60$ $m$ length. As illustrated in Fig. \ref{SCENARIO_PLOT}, the center of the square area is named the virtual SN center, and users are uniformly distributed within the area of $80$ $m$ radius. The rest of the simulation parameters are listed in Table \ref{table_sim}, where $R$ is the multimedia downloading rate required by each user, and $R/T$ is the size of the total traffic demand.

\begin{table}[ht]
\caption{Simulation Parameters}\label{table_sim} 
\centering\  
\begin{tabular}{|c|c|}
\hline
Carrier Frequency& $2000$ $MHz$\\
\hline
$B$ &$180$ $kHz$\\
\hline
$M$&$4$\\
\hline
$N$&$100$\\
\hline
$K$&$30$\\
\hline
Cell Radius& $250$ $m$ (eNB)\\
\hline
Path Loss Model& $\begin{array}{l}
L_{eNB} = 128.1 + 37.6log10 ( d/1000 )\\
L_{SN} =37+30log10(d)+18.3( 4^{\frac{4+2}{4+1}-0.46})
\end{array}$\\
\hline
Shadowing & $10$ $dB$ Log Normal Fading\\
\hline
$N_0$ & $-174$ $dBm/Hz$\\
\hline
Noise Figure &$5$ $dB$ (eNB/SN), $9$ $dB$ (UE)\\
\hline
Antenna Gain & $16$ $dB$ (eNB), $5$ $dB$ (SN)\\
\hline 
Fragment size& $S={R}/({TD})$\\
\hline
\end{tabular}
\label{tab_2}
\end{table}


\begin{figure}
\vspace{-1em}
        \centering   
                \begin{subfigure}[!t]{3 in}
                \includegraphics[width=3in,height=2.5in]{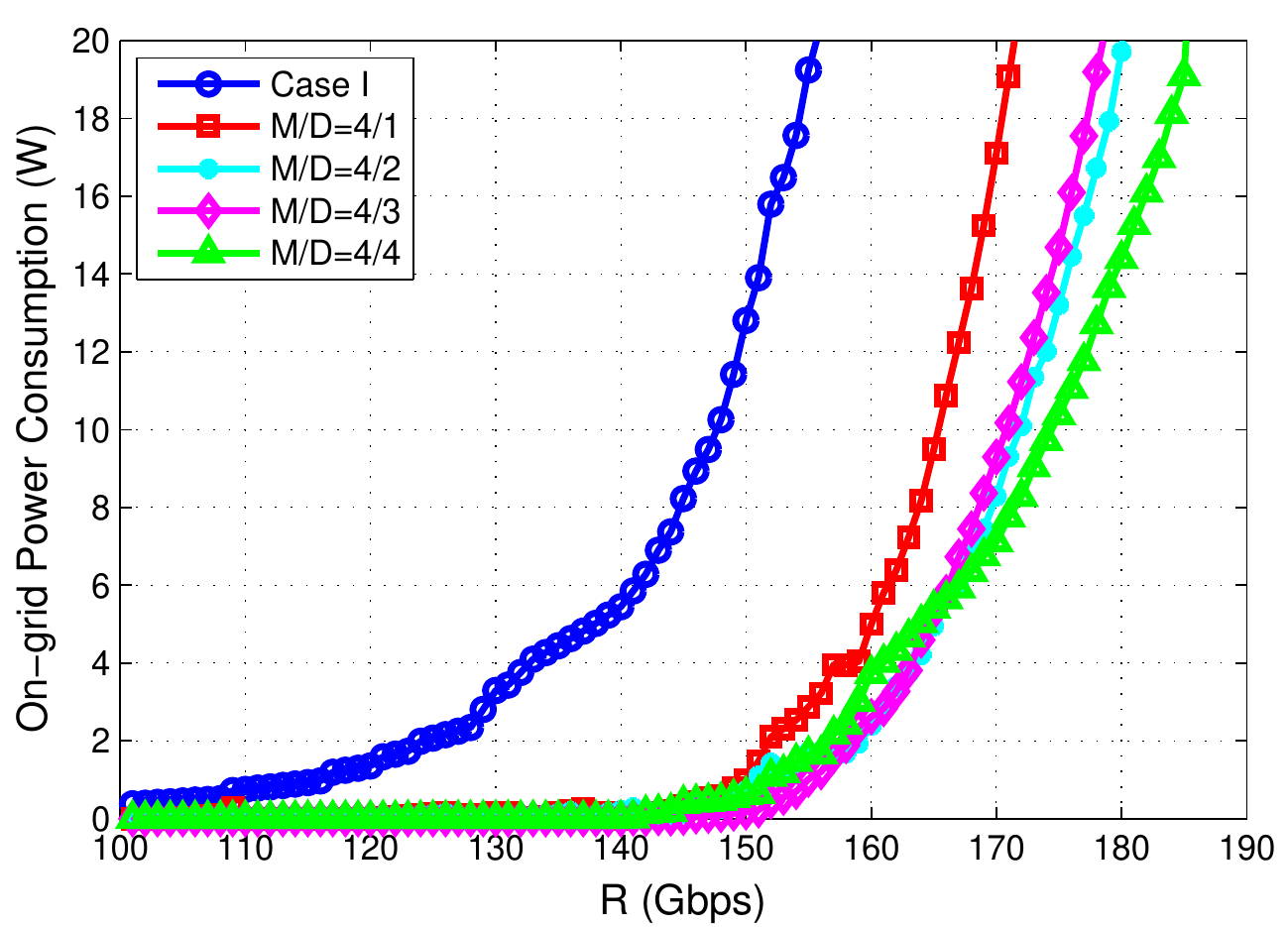}
                \caption{$\theta=0$}
        \end{subfigure}
        ~
        
        \begin{subfigure}[!t]{3 in}
                \includegraphics[width=3in,height=2.5in]{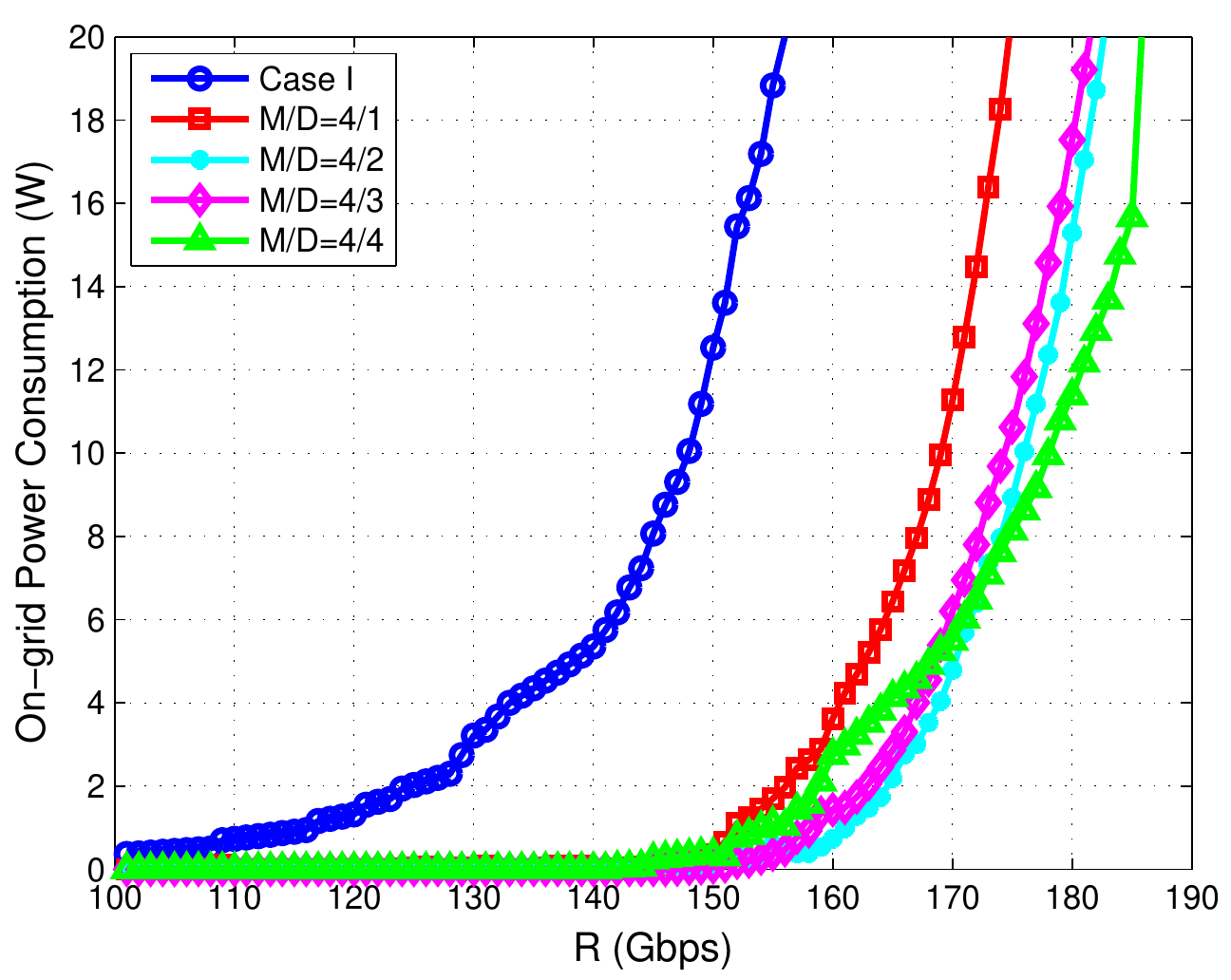}
                \caption{$\theta=0.4$}
        \end{subfigure}%
        ~
        
         ~
        \begin{subfigure}[!t]{3 in}
                \includegraphics[width=3in,height=2.5in]{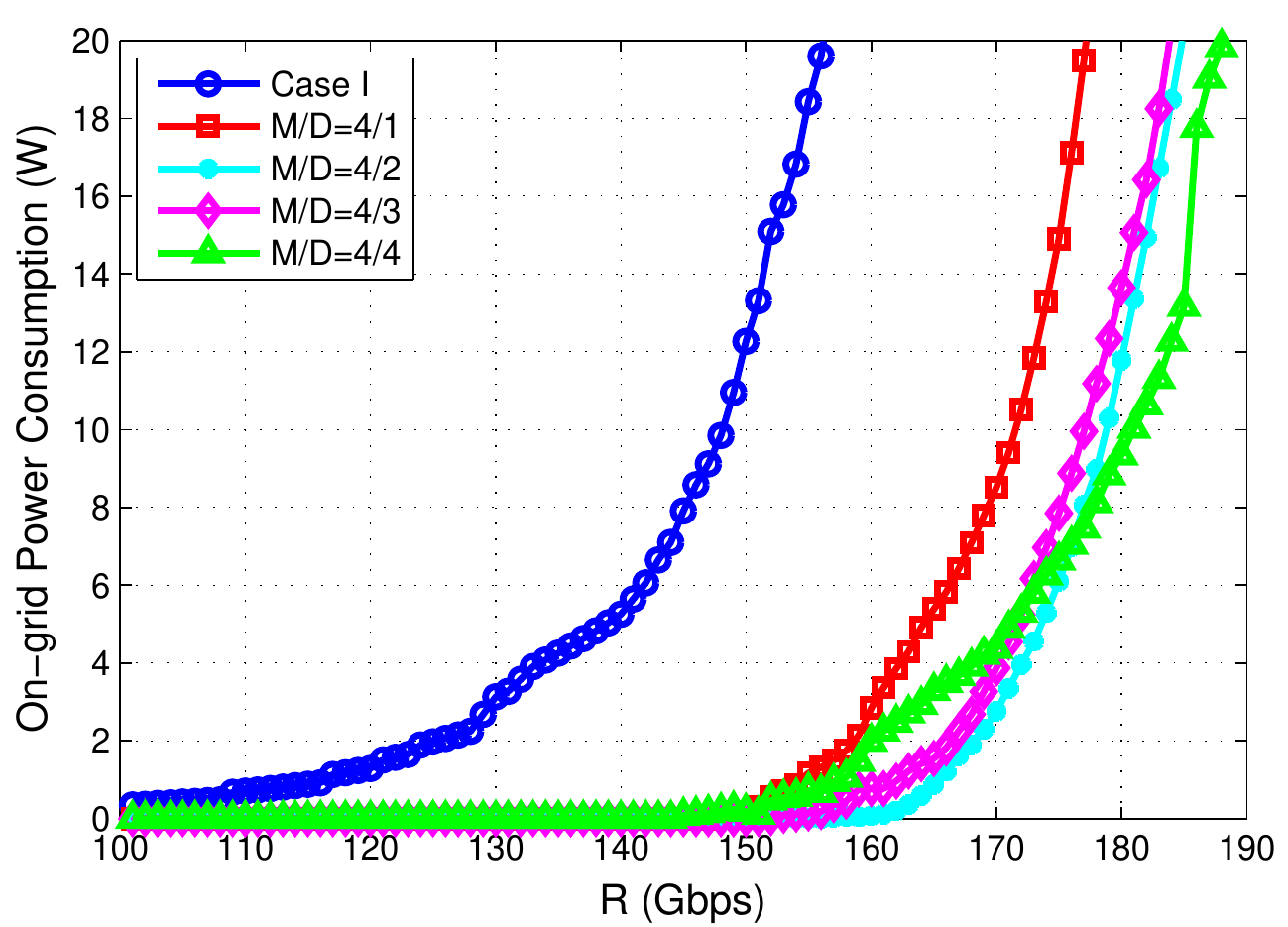}
               \caption{$\theta=0.8$}
        \end{subfigure}
        ~ 

        \caption{Total on-grid power consumption versus multimedia download rate (cell edge).}\label{result}
\end{figure}

The on-grid power consumption are averaged over $100$ independent snapshots by Monte-Carlo simulations. Using Case I, where only eNB is active, as the base line, we compare the total on-grid power consumption of our proposed algorithm versus the downloading rate $R$.

\subsection{Cell Edge Users}
First, we measure the performance of the virtual SN center located at the cell edge, i.e., the distance between the virtual SN center and eNB is at least $0.6$ times the radius of the cell. As illustrated in Fig. \ref{result},  the on-grid power consumption will reduce as $\theta$ increases. More importantly, the reduction of on-grid power consumption in Case I is less obvious than the proposed algorithm. This is because in setting the parameters, $E_1$ is significantly higher than the energy generating rate of SNs. So, the power flow from SN to eNB will be significantly less than the power flow from eNB to SNs, and thus the latter power flow is more prone to be affected by the energy transfer efficiency.

At the same time, the on-grid power consumption will increase with the data rate, and the proposed scheme outperforms the traditional multicast case. This shows that bringing serving nodes, cached content, and green energy closer to the cell edge users can save energy. Moreover, $D=1$ means the replication caching scheme. It has the highest effective redundancy factor $M/D$, and yet brings the least power reduction. The reason is that increased traffic load due to high redundancy trumps the diversity gain in terms of choosing different serving nodes. 

Diverse serving nodes provide the potential to increase the minimum SNR and reduce the power consumption. To leverage this potential, multiple SNs need to be activated. When $D=1$, the active SNs bring the most amount of traffic load, since $S$ in Table \ref{table_sim} will decrease with $D$. So, as analyzed in Section \ref{assump}, with a given number of subchannels, the increased traffic load can offset the channel diversity gain. 

The on-grid power, however, does not always decrease with $D$. For example, $D=4$ means no diversity gain, because users have to connect with all SNs, but it always has the minimum traffic load. With $D_3=3$ and $D_4=4$, the corresponding traffic size in each SN is $S_3$ and $S_4$, respectively. By the definition given in Table \ref{table_sim}, $S_3-S_4=(1/D_3-1/D_4){R}/T$ will increase with $R$. When $R$ is small, the traffic load size $S_3$ is not that large as compared to $S_4$, and so the diversity gain may cancel out the traffic load increment. When $R$ exceeds a certain point, the relatively decreased traffic load in $D_4=4$ will trump the diversity gain.

\subsection{Cell Center Users}
As we can see from Fig. \ref{result}, the maximum on-grid power is limited by $20$ $W$, and so the maximum transmission power of eNB and SN are nearly $40$ $W$. For the eNB in Case I, which is located at the cell center, this power limit complies with the current LTE standard. For the virtual SN center located at the cell edge, the maximum transmission power of each SN is between $10$ $W$ to $40$ $W$, in which unbearable interference may cause the performance degradation of the other virtual SN center which happens to locate at the edge of the neighboring cell. 

\begin{figure}
\vspace{-1em}
        \centering   
                \begin{subfigure}[!t]{3 in}
                \includegraphics[width=3in,height=2.5in]{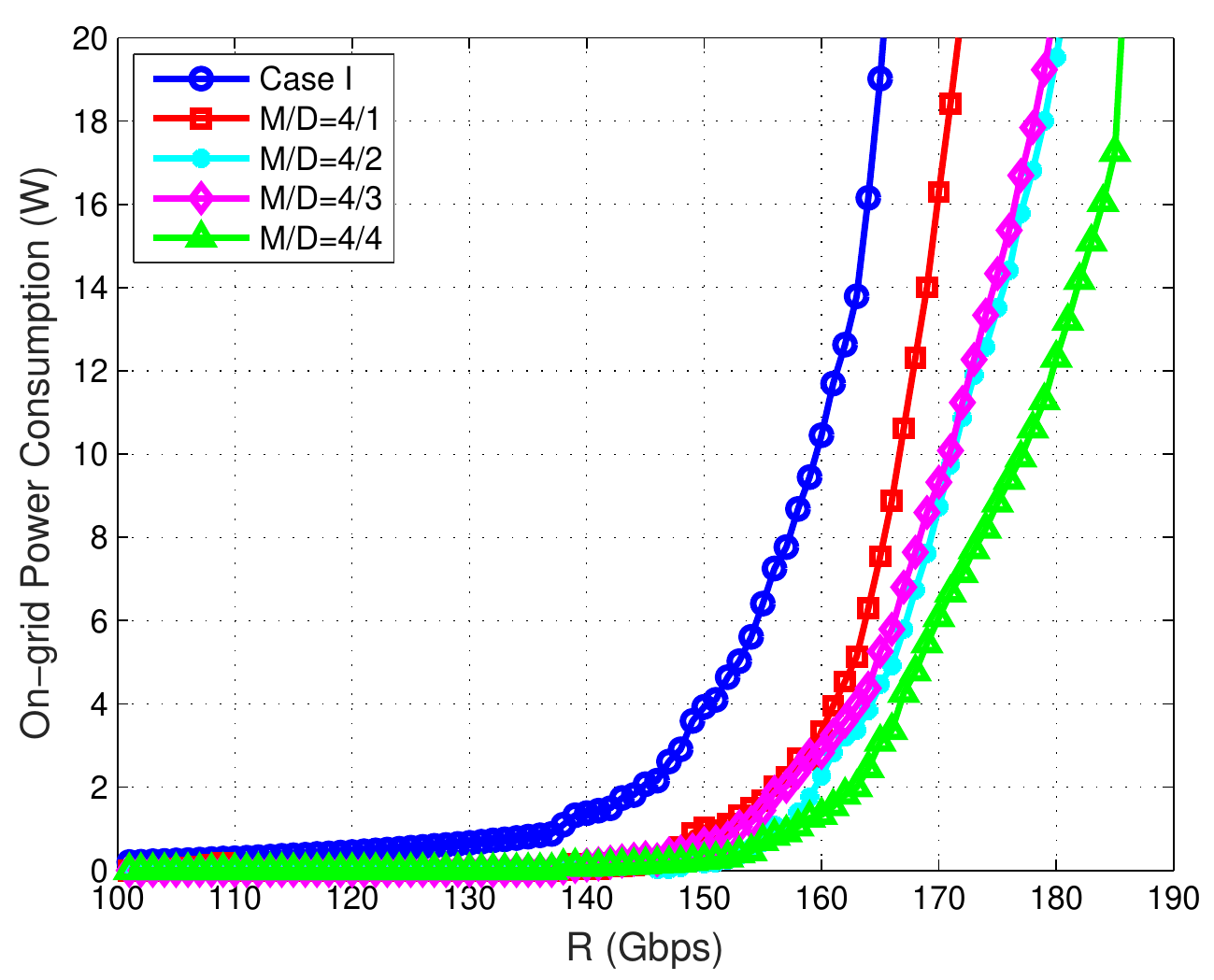}
                \caption{$\theta=0$}
        \end{subfigure}
        ~
        
        \begin{subfigure}[!t]{3 in}
                \includegraphics[width=3in,height=2.5in]{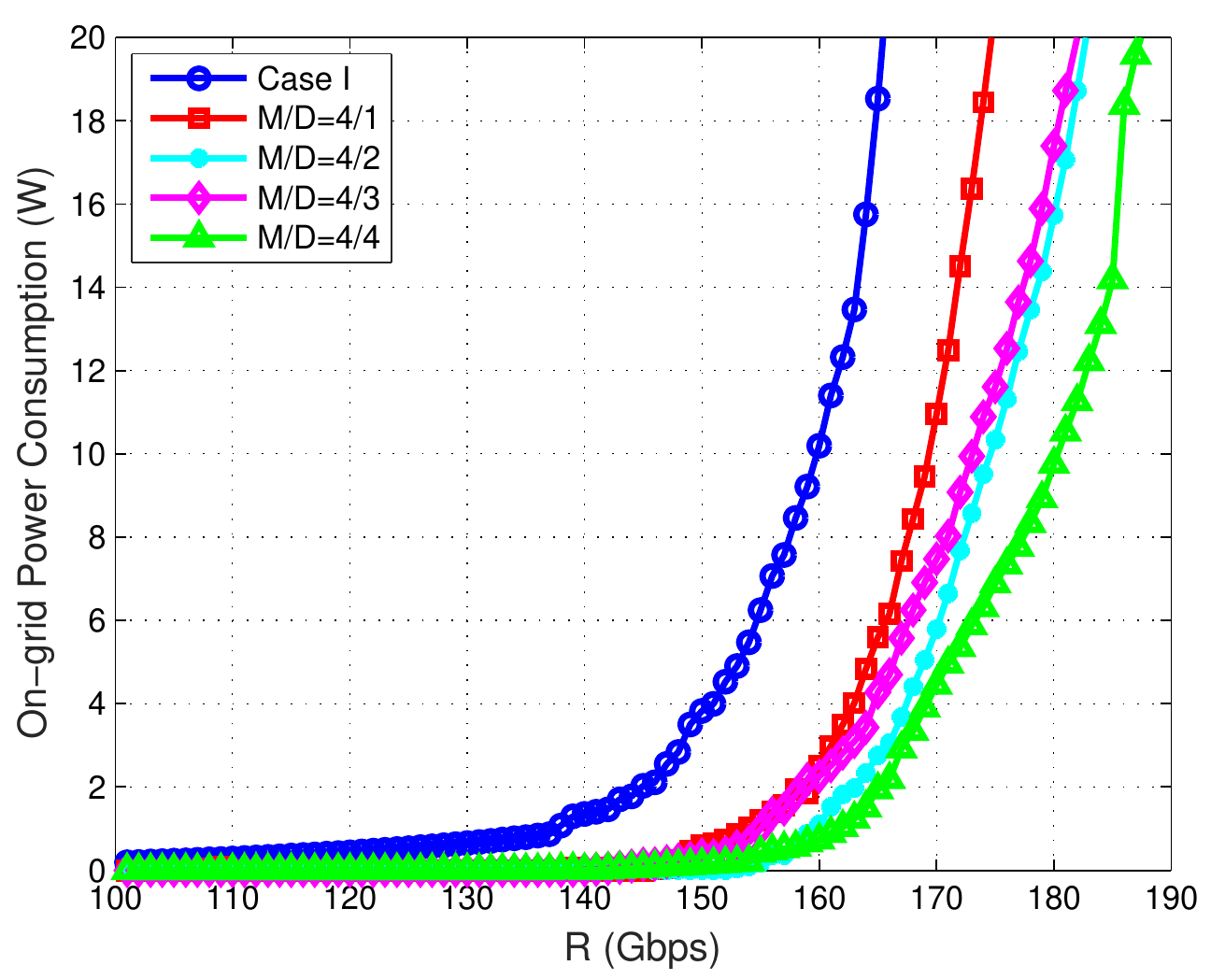}
                \caption{$\theta=0.4$}
        \end{subfigure}%
        ~
        
         ~
        \begin{subfigure}[!t]{3 in}
                \includegraphics[width=3in,height=2.5in]{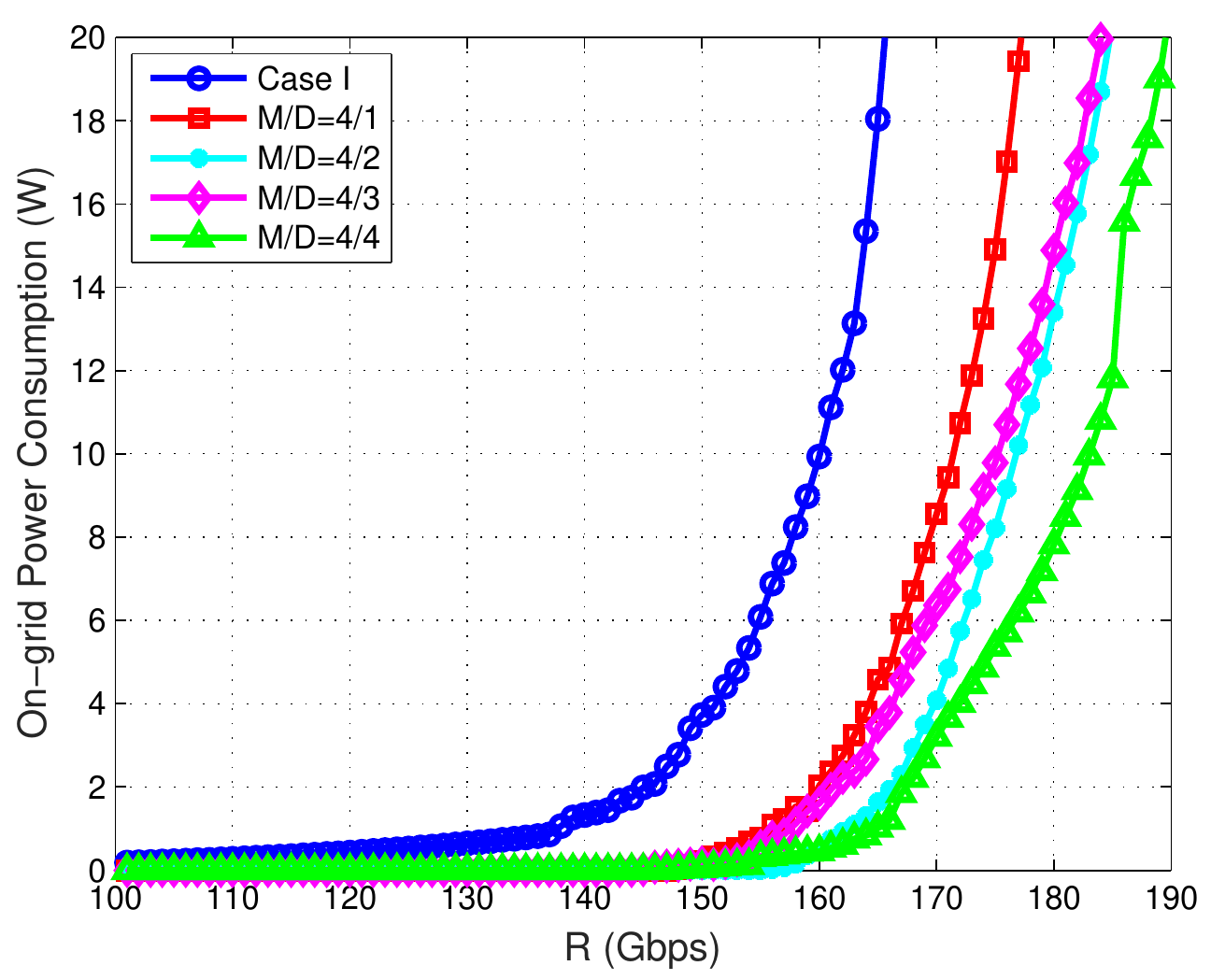}
               \caption{$\theta=0.8$}
        \end{subfigure}
        ~ 

        \caption{Total on-grid power consumption versus multimedia download rate (cell center).}\label{result2}
\end{figure}

So, in this section, we measure the performance for the virtual SN center located at the cell center, i.e., the distance between the virtual SN center and eNB is less than $0.6$ times the radius of the cell. As illustrated in Fig. \ref{result2}, the on-grid power consumption for Case I is less, as compared with the cell edge case. This is straight forward because users are closer to the eNB. The improvement for SNs are less obvious because the relative positions of SN and users remain the same. However, the proposed algorithm still outperforms Case I; this proves that bringing content, green energy and serving node closer to users can save the on-grid energy consumption. Moreover, the caching scheme with low effective redundancy factor (small $M/D$) can actually save more energy than high redundancy caching scheme when the data rate is high.


\section{Conclusion}
In this paper, we have proposed the content caching and distribution framework for the smart grid enabled wireless multimedia transmission system, where the green energy can be transferred to the serving node that is closer to end users through the grid credit exchange system. To investigate the on-grid energy reduction that is brought by closely located content, downlink node and green energy, the user association problem is designed jointly with radio resource allocation schemes, including power allocation, subchannel allocation, and power flow design. We have analytically derived the optimal power allocation and power flow scheme. Moreover, the user association problem has been successfully decoupled from the subchannel allocation problem. A novel maximal minimum SNR algorithm is proposed along with a two step subchannel assignment algorithm, which takes into account of the green energy generating rate of each node. Simulation results have proven that the proposed schemes outperform the traditional eNB multicasting scheme. That is, bringing content, serving node and green energy closer to the user can save the on-grid power consumption. Moreover, when the downloading rate requirement of each user is high, the caching scheme with low redundancy has been shown to save more energy than the high redundancy caching scheme.
\bibliographystyle{IEEEtran}
\bibliography{mybib}

\end{document}